\documentclass[journal]{IEEEtran}
\usepackage{amsmath,amsfonts}
\usepackage{amssymb}
\usepackage{algorithmicx,algorithm}
\usepackage{array}
\usepackage{textcomp}
\usepackage{stfloats}
\usepackage{url}
\usepackage{verbatim}
\usepackage{graphicx}
\usepackage{cite}
\usepackage{amsthm}
\usepackage{theorem}
\usepackage{subfigure}  
\usepackage{color}
\usepackage{multirow}

\begin{document}

\title{Joint Precoding Design and Resource Allocation for C-RAN Wireless Fronthaul Systems}

\author{Peng Jiang, Jiafei Fu, Pengcheng Zhu,~\IEEEmembership{Member,~IEEE,}, Jiamin Li,~\IEEEmembership{Member,~IEEE,} and Xiaohu You,~\IEEEmembership{Fellow,~IEEE,}
\thanks{This work was supported by the National Key R\&D Program of China under Grant 2021YFB2900300, and by the National Natural Science Foundation of China under Grant 62171126.\emph{(Corresponding author: Pengcheng Zhu.)}}
\thanks{The authors are with the National Mobile Communications Research
Laboratory, Southeast University, Nanjing 210096, China (e-mail: jiangpeng98@seu.edu.cn; fujfei@seu.edu.cn; p.zhu@seu.edu.cn; jiaminli@seu.edu.cn; xhyu@seu.edu.cn).}
}

\markboth{Journal of \LaTeX\ Class Files,~Vol.~14, No.~8, August~2021}%
{Shell \MakeLowercase{\textit{et al.}}: A Sample Article Using IEEEtran.cls for IEEE Journals}

\maketitle

\begin{abstract}
This paper investigates the resource allocation problem combined with fronthaul precoding and access link sparse precoding design in cloud radio access network (C-RAN) wireless fronthaul systems.
Multiple remote antenna units (RAUs) in C-RAN systems can collaborate in a cluster through centralized signal processing to realize distributed massive multiple-input and multiple-output (MIMO) systems and obtain performance gains such as spectrum efficiency and coverage.
Wireless fronthaul is a flexible, low-cost way to implement C-RAN systems, however, compared with the fiber fronthaul network, the capacity of wireless fronthaul is extremely limited.
Based on this problem, this paper first design the fronthaul and access link precoding to make the fronthaul capacity of RAUs match the access link demand.
Then, combined with the precoding design problem, the allocation optimization of orthogonal resources is studied to further optimize the resource allocation between fronthaul link and access link to improve the performance of the system.
Numerical results verify the effectiveness of the proposed precoding design and resource allocation optimization algorithm.

\end{abstract}


\section{Introduction}\label{Sec1}
\IEEEPARstart{C}{loud} radio access network (C-RAN) is a promising candidate for the next-generation communication system \cite{You2020SCIS}.
In the C-RAN system, most of the signal processing tasks are transferred to the cloud, known as the baseband unit (BBU), and connected to the remote antenna units (RAUs) with limited processing capacity via fronthaul links \cite{Antonioli2022TWC}.
Through centralized signal processing in the cloud, BBU can jointly design signals, precoding, or resource allocation, so that RAU can cooperate in different degrees to jointly serve users or suppress interference.
When RAUs are fully cooperative, the system can be equivalent to the massive multiple-input and multiple-output (MIMO) system composed of multiple RAUs, and obtain the gain that is similar to the massive MIMO system.
In the case of a partial cooperative system, some performance such as interference suppression can be achieved through the design of precoding, etc., to reduce the complexity of system implementation at the cost of system performance reduction.

Typical C-RAN systems, which assume fiber for fronthaul links, have high system capacity and enable full RAU cooperation for optimal system performance.
However, high laying cost and low flexibility of fixed fiber make the realization of the system more complicated.
In wireless fronthaul systems, the BBU equipped with multiple antennas communicates with RAU through the wireless link to transmit data. 
This system has high flexibility and low implementation difficulty.
However, compared with fiber links, the capacity of wireless links are relatively limited, which makes the fronthaul capacity limit become a serious problem that wireless fronthaul system has to consider. 

Existing studies on wireless fronthaul systems mostly focus on relay mode or only consider single antenna RAU to simplify the analysis.
Meanwhile, there are few studies on resource allocation optimization.
\cite{Park2017LWC} solves the problem of joint optimization of fronthaul link and access links across multiple clusters. 
However, it designs only single-antenna RAU, which has a great impact on the performance of fronthaul links.
\cite{Zhang2018TCOM} designed precoding for the combination of single-stream multi-antenna RAU and access link, but did not consider the case of multi-stream.
Survey of \cite{Tezergil2022COMST} on wireless fronthaul systems shows that the resource problem in wireless fronthaul systems is still an emerging problem that has not been fully studied. 
A few studies such as \cite{Muhammed2020TCOM,Zhang2018TCOM} consider the resource allocation problem. 
However, specific precoding design issues are not considered.

This paper studies the joint optimization of precoding design and resource allocation for wireless fronthaul systems.
Firstly, an MU-MIMO structure fronthaul link is employed and each RAU receives multiple data streams from the BBU to improve the performance of the fronthaul link with the total fronthaul and access link power limitation instead of separately limiting their power. 
Then, using semidefinite relaxation (SDR), successive convex approximation (SCA), and $l_0/l_1$ norm approximation method, this paper proposes an algorithm for joint optimization of fronthaul link and access link precoding.
Subsequently, a heuristic resource allocation method is proposed based on the relationship between fronthaul and access link, combined with power allocation and orthogonal resource allocation method, which can effectively improve the system sum rate.
Finally, simulation results verify the effectiveness of the proposed algorithm.

\section{System Model}\label{Sec2}
In the C-RAN wireless fronthaul system, $K$ single-antenna users are served by $N$ RAUs equipped with $N_{\mathrm{r}}$ antennas which are connected to the central BBU equipped with $N_{\mathrm{B}}$ antennas through wireless links.
In order to avoid interference between fronthaul and access link, we assume that fronthaul and access link transmission are completed on different orthogonal resource blocks, and the BBU communicates with all RAUs in the resource of $\tau_{\mathrm{R}}$, while RAUs and users communicate in the resource of $\tau_{\mathrm{U}}=1-\tau_{\mathrm{R}}$. 
$\tau_{\mathrm{R}}$ and $\tau_{\mathrm{U}}$ are resource allocation ratios.

\subsection{Fronthaul links}
Set the downlink channels from BBU to RAUs as $\mathbf{G}=[\mathbf{G}^{\mathrm{T}}_{1},\mathbf{G}^{\mathrm{T}}_{2},\ldots,\mathbf{G}^{\mathrm{T}}_{N}]^{\mathrm{T}}\in\mathbb{C}^{N_{\mathrm{R}} \times (N_{\mathrm{B}})}$, where $N_{\mathrm{R}}=N \times N_{\mathrm{r}}$.
The precoding matrix is $\mathbf{P}=[\mathbf{P}_{1},\mathbf{P}_{2},\ldots,\mathbf{P}_{N}]\in\mathbb{C}^{(N_{\mathrm{B}}) \times N_{\mathrm{R}}}$, and satisfies $\|\mathbf{P}\| \leq P_{\mathrm{BBU}}$.
Since RAUs share no information with each other, the receive matrix set as $\mathbf{Q}=\mathrm{diag}(\mathbf{Q}_1,\mathbf{Q}_2,\ldots,\mathbf{Q}_N)\in\mathbb{C}^{N_{\mathrm{R}} \times N_{\mathrm{R}}}$, which is a block diagonal matrix, where $\mathbf{Q}_n \in \mathbb{C}^{N_{\mathrm{r}} \times N_{\mathrm{r}}}$.
The noise vector is $\mathbf{n}=[n_1,n_2,\ldots,n_{N_{\mathrm{R}}}]\in\mathbb{C}^{N_{\mathrm{R}} \times 1}$, data stream vevtor is $\mathbf{s}\in\mathbb{C}^{N_{\mathrm{R}} \times 1}$.
We first introduce the following notations:
\begin{subequations}
    \begin{align}
        \mathbf{y}_{\mathrm{R},n}&=\left [ y_{\mathrm{R},n,1}, y_{\mathrm{R},n,2}, \ldots, y_{\mathrm{R},n,N_{\mathrm{r}}} \right ] \\
        \mathbf{Q}_{n}&=\left [ \mathbf{q}_{n,1}^{\mathrm{T}},\mathbf{q}_{n,2}^{\mathrm{T}}, \ldots, \mathbf{q}_{n,N_{\mathrm{r}}}^{\mathrm{T}}\right ] ^{\mathrm{T}}\\
        \mathbf{P}_{n}&=\left [ \mathbf{p}_{n,1}, \mathbf{p}_{n,2}, \ldots, \mathbf{p}_{n,N_{\mathrm{r}}} \right ]\\
        \mathbf{s}&=\left [ \mathbf{s}_{1}^{\mathrm{T}},\mathbf{s}_{2}^{\mathrm{T}}, \ldots, \mathbf{s}_{N}^{\mathrm{T}} \right ]^{\mathrm{T}}\\
        \mathbf{s}_{n}&=\left [ s_{n,1},s_{n,2}, \ldots, s_{n,N_\mathrm{r}} \right ]^{\mathrm{T}}.
    \end{align}
\end{subequations}
Assuming one receive antenna for one data stream, then for $i$-th stream, which is the $m$-th stream of RAU $n$ and $i=(n-1)N_r+m$, we have
\begin{equation}
    y_{\mathrm{R},i}=\mathbf{q}_{i} \mathbf{G}_n \mathbf{p}_{i}s_{i} + \sum_{t \neq i}^{N_{\mathrm{R}}} \mathbf{q}_{i} \mathbf{G}_n \mathbf{p}_{t}s_{t} + \mathbf{q}_{i}n_n.
\end{equation}
So the rate could be expressed as 
\begin{equation}\label{RateRi}
    C_{i}=\log_2 \left ( 1 + \frac{\mathbf{p}_{i}^{\mathrm{H}} \mathbf{G}_n^{\mathrm{H}} \mathbf{q}_{i}^{\mathrm{H}} \mathbf{q}_{i} \mathbf{G}_n \mathbf{p}_{i}}{\sum_{t \neq i}^{N_{\mathrm{R}}} \mathbf{p}_{t}^{\mathrm{H}} \mathbf{G}_n^{\mathrm{H}} \mathbf{q}_{i}^{\mathrm{H}} \mathbf{q}_{i} \mathbf{G}_n \mathbf{p}_{t}+\mathbf{q}_{i}\mathbf{q}^{\mathrm{H}}_{i}\sigma^2_{i}} \right ),
\end{equation}
So the fronthaul capacity of RAU $n$ is
\begin{equation}
    C_{\mathrm{FH},n} = \sum_{i\in\mathcal{N}_{R,n}}C_{i},
\end{equation}
where $\mathcal{N}_{R,n}=\{(n-1)N_r+1,(n-1)N_r+1+2,\ldots,n N_r\}$ is the set of antennas belong to RAU $n$.

Since the fronthaul link does not occupy all resources, the fronthaul capacity must be modified to $\tau_{\mathrm{R}} C_{\mathrm{FH},n}$.

\subsection{Access links for users}
Define the access link channel matrix from RAUs to users as $\mathbf{H}=[\mathbf{h}^T_1,\mathbf{h}^T_2,\ldots,\mathbf{h}^T_K ]^T \in \mathbb{C}^{K \times N_{\mathrm{R}}}$, and the precoding matrix as $\mathbf{W}=\left [\mathbf{w}_1,\mathbf{w}_2,\ldots,\mathbf{w}_K \right ] \in \mathbb{C}^{N_{\mathrm{R}} \times K}$ which satisfies $\|\mathbf{W}\|^2_F \leq P_{\mathrm{R}}$.
Then the receive signal vector $\mathbf{y}_{\mathrm{U},k}$ for use $k$ could be expressed as
\begin{equation}
    \mathbf{y}_{\mathrm{U},k}=\mathbf{h}_{k} \mathbf{w}_{k} x_{k} + \sum_{i \neq k}^K \mathbf{h}_{k}\mathbf{w}_{i} x_{i} + n_{k},
\end{equation}
Then the achievable rate could be written as
\begin{equation} \label{Rate_U}
    R_{k}=\log_2\left( 1 + \frac{\mathbf{w}^{\mathrm{H}}_{k}\mathbf{h}^{\mathrm{H}}_{k}\mathbf{h}_{k}\mathbf{w}_{k}}{\sum_{i \neq k}^K \mathbf{w}^{\mathrm{H}}_{i}\mathbf{h}^{\mathrm{H}}_{k}\mathbf{h}_{k}\mathbf{w}_{i}+\sigma^{2}}\right).
\end{equation}

Furthermore, in order to study the transmission from RAUs to users, the precoding matrix can be written as:
\begin{equation}
    \mathbf{W} =
    \left [ 
    \begin{array}{ccc}
    {{{\mathbf{w}}_{1,1}}}& \ldots &{{{\mathbf{w}}_{1,K}}}\\
     \vdots & \ddots & \vdots \\
    {{{\mathbf{w}}_{N,1}}}& \cdots &{{{\mathbf{w}}_{N,K}}}
    \end{array}
    \right ].
\end{equation}
where $\mathbf{w}_{n,k} \in \mathbb{C}^{N_{\mathrm{r}} \times 1}$.
In this case, we can use $l_0$-norm to check the link between RAUs and users.
$\|\|\mathbf{w}_{n,k}\|^2_F\|_{l_0}=1$ shows $\mathbf{w}_{n,k}$ is nonzero, which means RAU $n$ is serving user $k$.
Otherwise, when $\|\|\mathbf{w}_{n,k}\|^2_F\|_{l_0}=0$ means $n$-th RAU does serve user $k$.
Therefore, traffic undertaken by RAU $n$ can be expressed as follows:
\begin{equation}
    \tau_{\mathrm{U}}\sum^K_{k} \|\|\mathbf{w}_{n,k}\|^2_F\|_{l_0} R_{k} \leq \tau_{\mathrm{R}} C_{\mathrm{FH},n}.
\end{equation}

\section{Joint optimization of resource allocation and precoding for wireless fronthaul systems}\label{Sec3}
\subsection{Problem Formulation}
The optimization problem can be stated as
\begin{subequations}\label{MAXrate}
    \begin{align}
      \mathcal{P}_0: \mathop {\max }\limits_{\mathbf{P},\mathbf{Q},\mathbf{W},\tau_{\mathrm{U}},\tau_{\mathrm{R}}} \quad
        & \tau_{\mathrm{U}} \sum^K_{k} R_{k} \\
      {\rm{s.t.}} \quad 
        &\tau_{\mathrm{R}}\|\mathbf{P}\|^2_{F} + \tau_{\mathrm{U}}\|\mathbf{W}\|^2_{F} \leq P \label{stPow}\\
        &\tau_{\mathrm{U}}\sum^K_{k} \|\|\mathbf{w}_{n,k}\|^2_F\|_{l_0} R_{k} \leq \tau_{\mathrm{R}} C_{\mathrm{FH},n}\label{stCap}.
    \end{align}
\end{subequations}
Both the objective function and the constraint conditions are non-convex or involve multiplication problem, which is difficult to solve directly.
Therefore, we treated precoding optimization and resource allocation optimization separately and then merged them.

\subsection{Joint optimization of precodings for wireless fronthaul systems}
In this section, we fix resource allocation and only optimize the precoding of fronthaul link and access link.

\subsubsection{Reweighted $l_1$-norm approximation}
For the access link sparse precoding problem, we can use the convex function reweight $l_1$ norm to approximate the non-convex $l_0$ norm \cite{Dai2016JSAC}, i.e
\begin{equation}
     \|\|\mathbf{w}_{n,k}\|^2_F\|_{l_0} \approx \beta_{n,k}\|\mathbf{w}_{n,k}\|^2_F,
\end{equation}
where $\beta_{n,k}$ is the weight, which could be updated by SCA:
\begin{equation}\label{beta_update}
    \beta^{t+1}_{n,k}=\frac{1}{\|\mathbf{w}^{t}_{n,k}\|^2_{F}+\mu},
\end{equation}
where $\mu=10^{-5}$.
The weight of the $t+1$ iteration $\beta^{t+1}_{n,k}$ can be calculated from the result of the $t$ iteration.

\subsubsection{SDR for precoding vector}
We use SDR to deal with the precoding vector \cite{Xu2020TCOM}, set $\widetilde{\mathbf{P}}_i=\mathbf{p}_i \mathbf{p}^{\mathrm{H}}_i$, $\widetilde{\mathbf{Q}}_i=\mathbf{q}_i \mathbf{q}^{\mathrm{H}}_i$, $\widetilde{\mathbf{H}}_k=\mathbf{h}^{\mathrm{H}}_k \mathbf{h}_k$ and $\widetilde{\mathbf{W}}_k=\mathbf{w}_k \mathbf{w}^{\mathrm{H}}_k$.
Then Eq. \eqref{RateRi} can be rewritten as
\begin{equation}\label{RateRiSD}
    \begin{aligned}
    C_{i}&=\log_2 \left ( 1 + \frac{\mathrm{Tr}(\mathbf{p}_{i}^{\mathrm{H}} \mathbf{G}_n^{\mathrm{H}} \mathbf{q}_{i}^{\mathrm{H}} \mathbf{q}_{i} \mathbf{G}_n \mathbf{p}_{i})}{\sum_{l \neq i}^{N_{\mathrm{R}}} \mathrm{Tr}(\mathbf{p}_{l}^{\mathrm{H}} \mathbf{G}_n^{\mathrm{H}} \mathbf{q}_{i}^{\mathrm{H}} \mathbf{q}_{i} \mathbf{G}_n \mathbf{p}_{l})+\mathrm{Tr}(\mathbf{q}_{i}\mathbf{q}^{\mathrm{H}}_{i} \sigma^2_{i})} \right )\\
    &=\log_2 \left ( \sum_{l}^{N_{\mathrm{R}}} \mathrm{Tr}(\widetilde{\mathbf{P}}_l \mathbf{G}_n^{\mathrm{H}} \widetilde{\mathbf{Q}}_i \mathbf{G}_n )+\mathrm{Tr}(\widetilde{\mathbf{Q}}_i \sigma^2_{i}) \right )\\
    &-\log_2 \left ( \sum_{ l \neq i }^{N_{\mathrm{R}}} \mathrm{Tr}(\widetilde{\mathbf{P}}_l \mathbf{G}_n^{\mathrm{H}} \widetilde{\mathbf{Q}}_i \mathbf{G}_n )+\mathrm{Tr}(\widetilde{\mathbf{Q}}_i \sigma^2_{i}) \right ).
    \end{aligned}
\end{equation}
Eq. \eqref{Rate_U} can also be rewritten as
\begin{equation}
    \begin{aligned}
        R_{k}&=\log_2\left( 1 + \frac{\mathrm{Tr}(\mathbf{w}^{\mathrm{H}}_{k}\mathbf{h}^{\mathrm{H}}_{k}\mathbf{h}_{k}\mathbf{w}_{k})}{\sum_{i \neq k} \mathrm{Tr}(\mathbf{w}^{\mathrm{H}}_{i}\mathbf{h}^{\mathrm{H}}_{k}\mathbf{h}_{k}\mathbf{w}_{i})+\sigma^{2}}\right)\\
        &=\log_2 \left ( \sum_{i}\mathrm{Tr}(\widetilde{\mathbf{W}}_k\widetilde{\mathbf{H}}_k)+\sigma^{2}  \right )\\
        &-\log_2 \left ( \sum_{i \neq k}\mathrm{Tr}(\widetilde{\mathbf{W}}_k\widetilde{\mathbf{H}}_k)+\sigma^{2}  \right )
    \end{aligned}
\end{equation}

\subsubsection{SCA for difference-of-concave problem}
$C_{i}$ has a difference-of-concave item, this item can be approximated as an affine function by using Taylor expansion \cite{Guo2019ACCESS}. 
Set
\begin{subequations}
    \begin{align}
        C_{i}&=C_{\mathrm{Rci},i} - C_{\mathrm{Itf},i}\\
        C_{\mathrm{Rci},i}&=\log_2 \left ( \sum_{l}^{N_{\mathrm{R}}} \mathrm{Tr}(\widetilde{\mathbf{P}}_l \mathbf{G}_n^{\mathrm{H}} \widetilde{\mathbf{Q}}_i \mathbf{G}_n )+\mathrm{Tr}(\widetilde{\mathbf{Q}}_i \sigma^2_{i}) \right )\\
        C_{\mathrm{Itf},i}&=\log_2 \left ( \sum_{ l \neq i }^{N_{\mathrm{R}}} \mathrm{Tr}(\widetilde{\mathbf{P}}_l \mathbf{G}_n^{\mathrm{H}} \widetilde{\mathbf{Q}}_i \mathbf{G}_n )+\mathrm{Tr}(\widetilde{\mathbf{Q}}_i \sigma^2_{i}) \right ).
    \end{align}
\end{subequations}

Using first-order Taylor series expansions, the term $C_{\mathrm{Itf},i}$ can be approximated to an affine function of $\widetilde{\mathbf{P}}_l$.
This function in the iteration is constantly updated to ensure the accuracy of the approximation.
If $t$ times iteration has been carried out, then in the $(t+1)$-th iteration we have
\begin{equation}
    C_{\mathrm{Itf},i}^{t+1} \approx \widehat{C}_{\mathrm{Itf},i}^{t+1} = C_{\mathrm{Itf},i}^{t} + \sum_{l \neq i}^{N_{\mathrm{R}}} \langle \widetilde{\mathbf{P}}_l-\widetilde{\mathbf{P}}^{t}_l,\frac{\partial C^t_{\mathrm{Itf},i}}{\partial \widetilde{\mathbf{P}}^{t}_l}\rangle,
\end{equation}
where $\langle A,B \rangle=\mathrm{Tr}(AB^{\mathrm{H}})$ and
\begin{equation}
    \frac{\partial C_{\mathrm{Itf},i}}{\partial \widetilde{\mathbf{P}}_l} = 
    \frac{(\mathbf{G}_n^{\mathrm{H}} \widetilde{\mathbf{Q}}_i \mathbf{G}_n)^{\mathrm{H}}/\ln{2}}
    {\sum_{ l \neq i }^{N_{\mathrm{R}}} \mathrm{Tr}(\widetilde{\mathbf{P}}_l \mathbf{G}_n^{\mathrm{H}} \widetilde{\mathbf{Q}}_i \mathbf{G}_n )+\mathrm{Tr}(\widetilde{\mathbf{Q}}_i \sigma^2_{i})}.
\end{equation}
$\widetilde{\mathbf{Q}}_i$ is fixed during the optimization, and will be updated in the iteration. 
When MRC combination method is adopted, the receive combination vector can be updated by
\begin{subequations}
    \begin{align}\label{MRC}
    \mathbf{q}_i&=(\mathbf{G}_n \mathbf{p}_i)^{\mathrm{H}}\\
    \widetilde{\mathbf{Q}}_i&=\mathbf{q}^{\mathrm{H}}_i \mathbf{q}_i=\mathbf{G}_n \mathbf{p}_i \mathbf{p}^{\mathrm{H}}_i \mathbf{G}^{\mathrm{H}}_n.
    \end{align}
\end{subequations}

For the access link transmission, we also have
\begin{subequations}
    \begin{align}
        R_{k}&=R_{\mathrm{Rci},k} - R_{\mathrm{Itf},k}\\
        R_{\mathrm{Rci},k}&=\log_2 \left ( \sum_{l}\mathrm{Tr}(\widetilde{\mathbf{W}}_k\widetilde{\mathbf{H}}_k)+\sigma^{2}  \right )\\
        R_{\mathrm{Itf},k}&=\log_2 \left ( \sum_{l \neq k}\mathrm{Tr}(\widetilde{\mathbf{W}}_k\widetilde{\mathbf{H}}_k)+\sigma^{2}  \right ),
    \end{align}
\end{subequations}
then we get
\begin{subequations}
    \begin{align}
        \frac{\partial R_{\mathrm{Itf},k}}{\partial \widetilde{\mathbf{W}}_l} &= 
        \frac{\widetilde{\mathbf{H}}_k/\ln{2}}{\sum_{l \neq k}\mathrm{Tr}(\widetilde{\mathbf{W}}_k\widetilde{\mathbf{H}}_k)+\sigma^{2}}\\
        R_{\mathrm{Itf},k}^{t+1} &\approx \widehat{R}_{\mathrm{Itf},k}^{t+1} = R_{\mathrm{Itf},k}^{t} + \sum_{l \neq i}^{N_{\mathrm{R}}} \langle \widetilde{\mathbf{W}}_l-\widetilde{\mathbf{W}}^{t}_l,\frac{\partial R^t_{\mathrm{Itf},k}}{\partial \widetilde{\mathbf{W}}^{t}_l}\rangle .
    \end{align}
\end{subequations}

Then the optimization problem for joint precoding design can be approximated as an convex optimization problem
\begin{subequations}\label{CVX_PtW}
    \begin{align}
      &\mathcal{P}_1: \mathop {\max }\limits_{\widetilde{\mathbf{P}}^{t+1}_i,\widetilde{\mathbf{W}}^{t+1}_k} \quad
         \tau_{\mathrm{U}} \sum_{k \in \mathcal{K}_{n}} \widehat{R}^{t+1}_{k} \\
      &{\rm{s.t.}}\nonumber\\
        &\tau_{\mathrm{R}}\sum_i \mathrm{Tr}(\widetilde{\mathbf{P}}^{t+1}_i) + \tau_{\mathrm{U}}\sum_k \mathrm{Tr}(\widetilde{\mathbf{W}}^{t+1}_k) \leq P_{\mathrm{BBU}} \\
        &\tau_{\mathrm{U}}\sum_{k} \beta^{t+1}_{n,k} \mathrm{Tr}(\widetilde{\mathbf{W}}^{t+1}_{k|n}) \widehat{R}^t_{k} \leq \tau_{\mathrm{R}} \widehat{C}^t_{\mathrm{FH},n}\\
        &\widehat{R}^{t+1}_{k} \leq R^{t}_{\mathrm{Rci},k}- \left ( R_{\mathrm{Itf},k}^{t} + \sum_{l \neq i}^{N_{\mathrm{R}}} \langle \widetilde{\mathbf{W}}_l-\widetilde{\mathbf{W}}^{t}_l,\frac{\partial R^t_{\mathrm{Itf},k}}{\partial \widetilde{\mathbf{W}}^{t}_l}\rangle \right ) \\
        &\widehat{C}^{t+1}_{i} \leq C^t_{\mathrm{Rci},i}-\left ( C_{\mathrm{Itf},i}^{t} + \sum_{t \neq i}^{N_{\mathrm{R}}} \langle \widetilde{\mathbf{P}}_l-\widetilde{\mathbf{P}}^{t}_l,\frac{\partial C^t_{\mathrm{Itf},i}}{\partial \widetilde{\mathbf{P}}^{t}_l}\rangle \right ),
    \end{align}n 
\end{subequations}
where $\widetilde{\mathbf{W}}^{t+1}_{k|n} \in \mathbb{C}^{N_{\mathrm{r}} \times N_{\mathrm{r}}}$ is a square matrix consisted of $N_{\mathrm{r}}(n-1)+1$ to $N_{\mathrm{r}}(n-1)+N_{\mathrm{r}}$ rows and columns of $\widetilde{\mathbf{W}}^{t+1}_{k}$.

\subsection{Joint allocation for orthogonal resource and power}
The orthogonal resources such as time and frequency are not fully exchangeable with power. 
The optimal performance of the system cannot be achieved simply by precoding design and power allocation which are contained in precoding design.
The optimal precoding within a certain power range is relatively fixed.
Therefore, orthogonal resources and power are jointly allocated in this section to find a better way of allocation within a small error range.
First, set the power and normalized signal-to-interference-plus-noise-ratio (SINR) as:
\begin{subequations}
    \begin{align}
    \lambda_{\mathrm{U},k}&=\mathrm{Tr}(\mathbf{w}\mathbf{w}^{\mathrm{H}})\\
    \lambda_{\mathrm{R},i}&=\mathrm{Tr}(\mathbf{p}_{i} \mathbf{p}_{i}^{\mathrm{H}})\\
    \gamma_{\mathrm{U},k}&=\frac{\mathrm{Tr}(\mathbf{w}^{\mathrm{H}}_{k}\mathbf{h}^{\mathrm{H}}_{k}\mathbf{h}_{k}\mathbf{w}_{k})/\mathrm{Tr}(\mathbf{w}\mathbf{w}^{\mathrm{H}})}{\sum_{i \neq k} \mathrm{Tr}(\mathbf{w}^{\mathrm{H}}_{i}\mathbf{h}^{\mathrm{H}}_{k}\mathbf{h}_{k}\mathbf{w}_{i})+\sigma^{2}}\\
    \gamma_{\mathrm{R},i}&=\frac{\mathrm{Tr}(\mathbf{p}_{i}^{\mathrm{H}} \mathbf{G}_n^{\mathrm{H}} \mathbf{q}_{i}^{\mathrm{H}} \mathbf{q}_{i} \mathbf{G}_n \mathbf{p}_{i})/\mathrm{Tr}(\mathbf{p}_{i} \mathbf{p}_{i}^{\mathrm{H}})}{\sum_{l \neq i}^{N_{\mathrm{R}}} \mathrm{Tr}(\mathbf{p}_{l}^{\mathrm{H}} \mathbf{G}_n^{\mathrm{H}} \mathbf{q}_{i}^{\mathrm{H}} \mathbf{q}_{i} \mathbf{G}_n \mathbf{p}_{l})+\mathrm{Tr}(\mathbf{q}_{i}\mathbf{q}^{\mathrm{H}}_{i} \sigma^2_{i})}.
    \end{align}
\end{subequations}
In massive MIMO systems, the interference between users is often well suppressed, so we assume that SINR is relatively fixed. 
In this case, the rate can be abbreviated as:
\begin{subequations}
    \begin{align}
        R_k&=\log_2 \left (1+\lambda_{\mathrm{U},k}\gamma_{\mathrm{U},k} \right )\\
        C_i&=\log_2 \left (1+\lambda_{\mathrm{R},i}\gamma_{\mathrm{R},i} \right ).
    \end{align}
\end{subequations}

Joint allocation for orthogonal resource and power optimization will be applied  after the near-convergence of precoding optimization.
In this case, the power distribution is close to optimal, that is, the distribution of water filling power distribution. 
In view of the distributed MIMO system, SINR is usually not too low, so it can be assumed that all SINRs reach the water filling line.
According to \cite{Chi2017CRC}, the power distribution of water filling can be written as
\begin{subequations}\label{Power_reall}
    \begin{align}
        \lambda_{\mathrm{U},k}&=\frac{P_{\mathrm{U}}+\sum^{K}_t \frac{1}{\gamma_{\mathrm{U},t}}}{K}-\frac{1}{\gamma_{\mathrm{U},k}}\\
        \lambda_{\mathrm{R},i}&=\frac{P_{\mathrm{R}}+\sum^{N_{\mathrm{R}}}_t \frac{1}{\gamma_{\mathrm{R},t}}}{N_{\mathrm{R}}}-\frac{1}{\gamma_{\mathrm{R},i}}.
    \end{align}
\end{subequations}
Then the rate can be expressed as
\begin{subequations}\label{Power_Rate}
    \begin{align}
        R_k&=\log_2 \left ( \frac{\gamma_{\mathrm{U},k}}{K}\left (P_{\mathrm{U}}+\sum^{K}_t \frac{1}{\gamma_{\mathrm{U},t}} \right ) \right )\\
        C_i&=\log_2 \left ( \frac{\gamma_{\mathrm{R},i}}{N_{\mathrm{R}}}\left (P_{\mathrm{R}}+\sum^{K}_t \frac{1}{\gamma_{\mathrm{R},t}} \right ) \right ).
    \end{align}
\end{subequations}

For wireless fronthaul systems, the optimal resource allocation will make the fronthaul capacity match the access link rate, namely:
\begin{equation}
    \tau_{\mathrm{U}}\sum^K_{k} \|\|\mathbf{w}_{n,k}\|^2_F\|_{l_0} R_{k} = \tau_{\mathrm{R}} C_{\mathrm{FH},n}.
\end{equation}
We approximate this condition of multiple RAUs to one and replace $l_0$ norm by a logical operation:
\begin{equation}
    \tau_{\mathrm{U}}\sum^N_n\sum^K_{k} (\|\mathbf{w}_{n,k}\|^2_{F}>\epsilon) R_{k} = \tau_{\mathrm{R}}\sum^{N_\mathrm{R}}_t C_{t},
\end{equation}
where $\mu \ll \epsilon \ll P/(NK)$ is a small power threshold, which is used to determine whether the RAU serves the user, we can set $\epsilon=0.01$.

In this case, an additional equation can be added, namely:
\begin{equation}
    |\tau_{\mathrm{U}}-\tau_{\mathrm{R}}|/2+|P_{\mathrm{U}}-P_{\mathrm{R}}|/(2P)=\tau .
\end{equation}
The resource difference should be as small as possible to match the rate between the fronthaul and the access link, so $\tau$ should be as small as possible.

At this point, we have the following equations:
\begin{subequations}\label{Tau_opt}
    \begin{align}
        &\tau_{\mathrm{U}}+\tau_{\mathrm{R}}=1\\
        &\tau_{\mathrm{U}}P_{\mathrm{U}}+\tau_{\mathrm{R}}P_{\mathrm{R}}=P\\
        &|\tau_{\mathrm{U}}-\tau_{\mathrm{R}}|/2+|P_{\mathrm{U}}-P_{\mathrm{R}}|/(2P)=\tau\\
        &\tau_{\mathrm{U}}\sum^N_n\sum^K_{k} \|\|\mathbf{w}_{n,k}\|^2_F\|_{l_0} R_{k} = \tau_{\mathrm{R}}\sum^{N_\mathrm{R}}_t C_{t}.
    \end{align}
\end{subequations}
Notice that $R_{k}$ and $C_{t}$ can be calculated through Eq. \eqref{Power_Rate} by $P_{\mathrm{U}}$ and $P_{\mathrm{R}}$.
As long as the value of $\tau$ is determined, this four-element equations that can be solved.
We use MATLAB to get numerical solution of $\tau_{\mathrm{U}},\tau_{\mathrm{R}}, P_{\mathrm{U}}, P_{\mathrm{R}}$ or return the information of no real solution.

Since $0<\tau<1$, binary search can be implemented on $(0,1)$.
Through $L$ times of search, the smallest feasible numerical solution with a precision of $\frac{1}{2^L}$ can be determined.
$L$ can be set as $6$.

\subsection{Summary of Algorithm}
In order to reduce the complexity and improve the feasibility of the algorithm, we optimize the resource allocation only when the conver bngence is near.
First, we take the mean square error of fronthaul and access link rates as the standard of iterative convergence:
\begin{equation}\label{D_calcu}
    D=\frac{1}{N}\sum^N_n \left ( \tau_{\mathrm{R}}C_{\mathrm{FH},n}-\tau_{\mathrm{U}}\sum^K_{k} \|\|\mathbf{w}_{n,k}\|^2_F\|_{l_0} R_{k} \right )^2 .
\end{equation}
When $D<1$, we optimize resource allocation, when $D<0.1$, the algorithm can be assumed that has converged.
Then the algorithm can be summarized as algorithm \eqref{JPDRA}.
\begin{algorithm}
    \caption{Joint Precoding Design and Resource Allocation Algorithm (JPDRA)}\label{JPDRA}
    \begin{algorithmic}[1]
    \State{\textbf{Input:} $\mathbf{H}$, $\mathbf{G}$}
    \State{initialize $D=1000$, $t=1$ and $\boldsymbol{\psi}=[0]$}
    \State{initialize $\mathbf{P^t}$ by SLNR \cite{Sadek2007TWC}, $\mathbf{W^t}$ by MRT,
    $\beta^{t}_{n,k}$ by Eq. \eqref{beta_update}}
    \State{\textbf{while} $\mathrm{length}(\boldsymbol{\psi})<2$ or $\mathrm{variance}(\boldsymbol{\psi})>=1$}
    \begin{enumerate}
         \State{Solve problem \eqref{CVX_PtW} using CVX, obtain $\widetilde{\mathbf{P}}^{t+1}_i,\widetilde{\mathbf{W}}^{t+1}_k$}
         \State{Recover $\mathbf{W}^{t+1},\mathbf{P}^{t+1}$ from $\widetilde{\mathbf{P}}^{t+1}_i,\widetilde{\mathbf{W}}^{t+1}_k$ by SVD}
         \State{Calculate all the $R_k$ and $C_i$}
         \State{Update $\beta^{t}_{n,k}$ by Eq. \eqref{beta_update}, $D$ by Eq. \eqref{D_calcu}, $\widetilde{\mathbf{Q}}^{t+1}$ by Eq. \eqref{MRC}}
         \State{\textbf{if} $D<1$}
         \begin{enumerate}
            \State{Record result $\boldsymbol{\psi}=[\boldsymbol{\psi},\tau_{\mathrm{U}} \sum^K_{k} R_{k}]$} and reserve three latest values of $\boldsymbol{\psi}$
            \State{Solve equations \eqref{Tau_opt} and obtain $\tau_{\mathrm{U}},\tau_{\mathrm{R}},P_{\mathrm{U}},P_{\mathrm{R}}$}
            \State{Reallocate the power of $\widetilde{\mathbf{P}}^{t+1}_i,\widetilde{\mathbf{W}}^{t+1}_k$ accroding to $P_{\mathrm{U}},P_{\mathrm{R}}$ and Eq. \eqref{Power_reall}}
         \end{enumerate}
         \State{\textbf{end if}}
         \State{$t \leftarrow t+1$}
    \end{enumerate}
    \State{\textbf{end while}}
    \State{\textbf{Output:} $\tau_{\mathrm{U}},\tau_{\mathrm{R}},\mathbf{W^{t+1}},\mathbf{P^{t+1}}$ }
    \end{algorithmic}
\end{algorithm}
$\mathbf{\psi}$ is the record of latest values of converged $\tau_{\mathrm{U}} \sum^K_{k} R_{k}$.
The JPDRA is converged when multiple convergence values do not change anymore, that is, the variance of multiple convergence values is small, i.e., $\mathrm{variance}(\mathbf{\psi})<1$.

\section{Numerical results} \label{Sec4}
In this section, simulations are presented to verify the performance of our proposed algorithms.
Set $ N_\mathrm{B}=28, N=3, N_\mathrm{r}=8, K=7, f_c=30\mathrm{GHz}, P=100\mathrm{W}$.
The channel parameters consist of path loss $-30.18+20\log_{10}(d)+20\log_{10}(f_c) \mathrm{dB}$ and gaussian complex gain $\mathcal{CN}(0,\sigma^2)$.
The initial resource allocation ratio is $\tau_{\mathrm{U}}=0.5,\tau_{\mathrm{R}}=0.5$.
Cell radius $R=500\mathrm{m}$, BBU is located in the center of the cell, and RAUs are evenly distributed on the ring of $\frac{2}{3}R$.

\begin{figure}
    \centering
    \includegraphics[width=\linewidth]{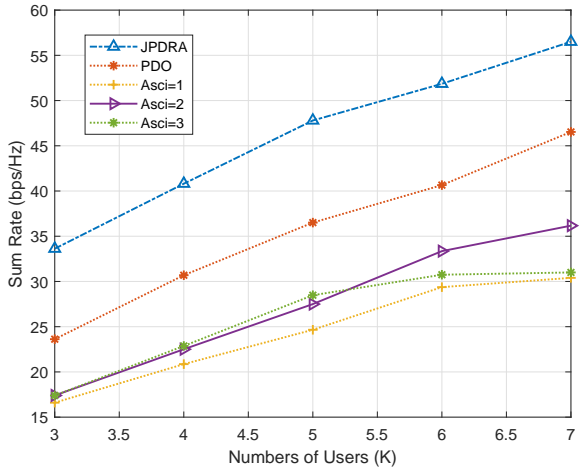}
    \caption{sum-rate versus the number of users}
    \label{COMP_K_TO}
\end{figure}

In Fig. \ref{COMP_K_TO}, JPDRA represents joint precoding design and resource allocation, PDO represents precoding design only. 
Asci represents fixed association method.
In this method, $\mathrm{Asci}= T_{\mathrm{Asci}}$ means one user is connected to $T_{\mathrm{Asci}}$ RAUs with the strongest channel gain, and MMSE precoding is used in access link.
From Fig. \ref{COMP_K_TO} we can see that PDO has obvious advantages over Asci method, while JPDRA further improve the performance compared to PDO.
These results verify the correctness of the proposed precoding optimization method and the necessity of resource allocation.

\begin{figure}
    \centering
    \includegraphics[width=\linewidth]{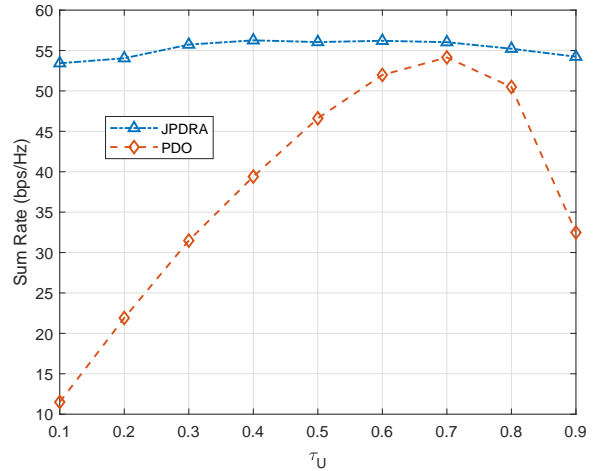}
    \caption{sum-rate versus $\tau_{U}$}
    \label{COMP_Tau_TO}
\end{figure}

According to Fig. \ref{COMP_Tau_TO}, we can see that the optimal resource allocation point under this system setting is around $\tau_{\mathrm{U}}=0.7,\tau_{\mathrm{R}}=0.3$.
In this resource configuration, the sum rate of JPDRA increases from $46.61\mathrm{bps/Hz}$ to $54.14\mathrm{bps/Hz}$ compared to the equal resource configuration $\tau_{\mathrm{U}}=0.5,\tau_{\mathrm{R}}=0.5$ of PDO. 
And compared to some of the more extreme resource configurations, such as $\tau_{\mathrm{U}} = 0.2, \tau_{\mathrm{R}} = 0.8 $, the performance gain of JPDRA is about $ 145.4\%$.
Therefore, the resource allocation between fronthaul links and access link is a very meaningful topic.
What's more, by observing the curves of JPDRA, it can be seen that the joint optimization method can optimize the allocation at any starting point and make it close to the optimal, which verifies the effectiveness and reliability of the proposed joint precoding design and resource allocation algorithm.

\section{Conclusion} \label{Sec7}
In this paper, the joint optimization of precoding and resource allocation in a C-RAN wireless fronthaul system is studied.
Using SDR, SCA, $l_0/l_1$ norm approximation and the convex optimization method, we solved the design problems of fronthaul and access link precoding.
Combined with the conclusions of the water-filling algorithm and the coupling relationship between fronthaul and access link, a joint optimization method of power and orthogonal resources allocation was designed.
According to the simulation results, the joint precoding design can effectively utilize the limited power and backhaul resources, while the orthogonal resource allocation enables the system to allocate resources more efficiently to achieve better performance.

\bibliographystyle{IEEEtran}

\bibliography{ref}

\end{document}